\documentclass[12pt,a4paper]{article}
\usepackage{amssymb}
\usepackage{amsmath}
\usepackage{float}
\usepackage{empheq}
\usepackage{cite}
\begin{document}

\begin{center}
{\Large \bf The specificity of the interactions of electroweak
gauge bosons coming from extra dimensions}
\end{center}

\vspace{4mm}

\begin{center}
{Eduard~Boos, Igor~Volobuev \\
\vspace{0.5cm}
{\small \it Skobeltsyn Institute of Nuclear Physics, Moscow State University}\\
{\small \it 119991, Moscow, Russia} }\\
\end{center}

\begin{abstract}
We discuss  the specificity of the interactions of the electroweak
gauge boson excitations in models with warped extra dimensions and
the Standard Model fields living in the bulk. In particular, we
show that the couplings of the gauge boson excitations
$W^{\prime}$, $Z^{\prime}$, and $\gamma^\prime$ to the SM gauge
bosons treated as the zero modes of the 5D gauge fields are either
exactly equal to zero or very much suppressed. In the former case,
the three-particle and four-particle interaction Lagrangians of
the SM gauge bosons and their lowest excitations are found
explicitly. Meanwhile, the couplings of $W^{\prime}$,
$Z^{\prime}$, and $\gamma^\prime$ to the SM fermions are non-zero
allowing for their production and decays. These are the
characteristic features of the gauge boson excitations in models
with warped extra dimensions, which distinguish them from the
gauge boson excitations in other models beyond the SM.
\end{abstract}

\section{Introduction}
Nowadays the Standard Model (SM) is capable of describing very
well a great amount of experimental facts and results. However,
there is a number of serious problems such as the hierarchy
problem,  dark matter, and  CP violation, which cannot be
consistently explained in its framework. To explain them, a large
number of various models  beyond the SM  and scenarios of new
physics have been put forward.  Almost all of the SM extensions
predict the existence of new particles, in particular, the
existence of massive charged and neutral vector particles besides
the gauge bosons of the SM. These extra vector bosons appear
either because of an extension of the SM gauge group (see e.g.,
\cite{Pati:1974yy,Kaplan:1983fs,Langacker:1984dc,Langacker:1989xa,
Chivukula:1995gu,Cvetic:1996mf,Batra:2003nj,Casalbuoni:2005rs}),
or as excitations of the SM gauge bosons (see e.g.,
\cite{Baur:1985ez,Datta:2000gm,Sundrum:2005jf,Csaki:2005vy,Kribs:2006mq,Agashe:2007ki,Agashe:2008jb,Alves:2009aa,deBlas:2012qf}).
The lowest excitations of $W$, $Z$, and $\gamma$ are usually
denoted by $W^{\prime}$, $Z^{\prime}$, and $\gamma^\prime$ or
para-photon. Depending on a particular model, the physical
properties and interactions of these extra vector bosons may be
rather different.

If extra vector bosons are found at the LHC, there arises the
problem of specifying the theory beyond the SM, to which they may
belong. To solve this problem one has to study the characteristic
features of the extra vector bosons in different models and the
specificity of their production channels and decay modes.

In the present paper we pay closer attention to the interactions
of the excitations of the electroweak SM gauge bosons in various
models with extra dimensions. Such models have been widely
discussed in the literature in the brane-world set-up either in
the context of the flat bulk (UED models)
\cite{Appelquist:2000nn,Rizzo:2001sd,Macesanu:2002ew,Belanger:2012mc},
or in the Randall-Sundrum bulk
\cite{Randall:1999ee,Pomarol:1999ad,
Davoudiasl:1999tf,Chang:1999nh,
Gherghetta:2000qt,Djouadi:2007eg,Allanach:2009vz, Agashe:2006wa,
Agashe:2007zd, Fitzpatrick:2007qr, Lillie:2007yh, Agashe:2006hk,
Burdman:2008qh, Agashe:2013kyb}. In fact, in the brane world
models there appear excitations of two neutral gauge bosons
$Z^{\prime}$, and $\gamma^\prime$, while in many BSM models based
on extensions of the SM gauge group only additional $Z^{\prime}$
bosons appear. This property is already a characteristic feature
of extra dimensional models. However the feature might not be very
pronounced since a typical mass splitting between $Z^{\prime}$ and
$\gamma^\prime$, as it will be discussed below, is expected to be
very small of the order of ${m_Z^2}/(2 {m_{\gamma^{\prime}}})$.
Such small mass splitting might be very difficult to resolve
experimentally.

It is a common knowledge that, in models with flat universal extra
dimensions, there exists the so-called Kaluza-Klein number
conservation, which is a trivial consequence of the properties of
the Fourier transform on the circle reflecting  the law of
momentum conservation in the extra dimension. It means that in
such models there is no single production at tree level of the
Kaluza-Klein excitations, and a Kaluza-Klein excitation cannot
decay at tree level  into the SM particles. Nevertheless, such
processes can take place due to loop corrections
\cite{Rizzo:2001sd}, which are usually very much suppressed. This
well-known property leads to rather specific collider signatures
with cascade decays and stable lightest state similar to SUSY
signatures but with different spin correlations in the decay
chains \cite{Datta:2010us}.

The brane-world models with the Randall-Sundrum bulk are rather
different  from the UED models with the flat bulk
\cite{Appelquist:2000nn, Rizzo:2001sd,
Macesanu:2002ew,Belanger:2012mc}, because the fields of different
tensor type have different wave function profiles. For this
reason, this scenario does not necessarily lead to the KK number
conservation. In this case the production of single KK states and
their decays are possible if kinematically allowed.

The extra vector bosons usually have interactions similar  to
those of the SM gauge bosons and can mediate the same processes
with SM particles. In this case non-trivial interference between
the contributions, for example, of $W$ and $W^{\prime}$, to
various processes \cite{Boos:2006xe,
Rizzo:2007xs,Accomando:2008jh,Accomando:2010ir,Accomando:2011xi,Accomando:2011eu}
could influence the experimental observation of the latter and the
exclusion limits  for their masses \cite{Abazov:2011xs,
Chatrchyan:2014koa, Aad:2014xea, Sirunyan:2018lbg}. Interference
of  $W^{\prime}$, $Z^{\prime}$ and $\gamma^{\prime}$ with the SM
bosons and with the rest of corresponding KK-towers coming from
extra dimensions leads to certain changes in invariant mass and
$P_t$ distributions \cite{Boos:2011ib,Boos:2013cxa}. However this
specific feature of the EW gauge boson KK excitations is also
rather delicate for experimental detection.

In the present paper we show that the interaction properties  of
EW gauge  boson  excitations in models with warped extra
dimensions are essentially different from the decay properties of
these excitations in other models. The interaction properties  of
the  excitations of the neutral EW gauge  bosons and gluons in the
unstabilized Randall-Sundrum model \cite{Randall:1999ee} have been
already touched upon in papers
\cite{Djouadi:2007eg,Allanach:2009vz}, the greater emphasis having
been put on the properties of the gluon excitations. Here we will
study the interaction properties of EW gauge  boson  excitations
in more detail  taking into consideration the excitations of the
charged SM gauge bosons. It will be demonstrated explicitly that a
simple common property of the EW gauge bosons excitations
$W^{\prime}$, $Z^{\prime}$, and $\gamma^\prime$  in brane-world
models is that their couplings to the SM gauge fields treated as
the  zero modes of the 5D gauge fields  are either exactly equal
to zero or very much suppressed. At the same time the couplings of
$W^{\prime}$, $Z^{\prime}$, and $\gamma^\prime$ to the SM fermions
are non-zero allowing for their production and decays.

\section{Electroweak gauge fields in the bulk}
Without loss of generality we consider a  brane-world RS model
stabilized by the bulk scalar field
\cite{Goldberger:1999uk,DeWolfe:1999cp} with the SM fields living
in the bulk. Such stabilized brane-world models can solve the
hierarchy problem of the gravitational interaction and give rise
to the masses of KK excitations of the SM fields and gravity field
in the $\textrm{TeV}$ energy range \cite{Boos:2005dc,Boos:2004uc}.
It is worth to mention that in the considered brane-world RS
models there are no flavour changing neutral currents strongly
suppressed by the present-day experimental data, because the
neutral currents have the same diagonal structure as in the SM.

Let us consider the electroweak gauge fields in a
five-dimensional space-time with the coordinates
$x^{M}=\{x^{\mu},y\}$, $M=0,1,2,3,4$. The compact extra dimension
forms the orbifold $ S^1/Z_2$, which can be represented as the
circle of circumference $2L$ with the coordinate $-L\le y\le L$
and the points $-y$ and $y$ identified. The background metric is
assumed to have the standard form, which is often used in brane
world models:
\begin{equation}\label{backgmetric}
ds^2 = \gamma_{MN}(y)dx^{M}dx^{N} =
e^{2\sigma(y)}\eta_{\mu\nu}dx^{\mu}dx^{\nu}-dy^2.
\end{equation}
This metric gives rise to a usual brane-world model, i.e. it is a
solution to the equations of motion for five-dimensional gravity,
two branes with tension and the stabilizing bulk scalar field. The
explicit form of the  solution for $\sigma(y)$ is unimportant for
our considerations and we do not specify it.

We consider the following standard action of the $SU(2)\times
U(1)$ gauge fields in this background:
\begin{equation}\label{action}
S= \frac{1}{{2L}}\int
d^{4}xdy\sqrt{\gamma}\left(-\frac{1}{4}W^{i,MN}W^{i}_{MN}-\frac{1}{4}B^{MN}B_{MN}\right),
\end{equation}
where  $\gamma =  \det\gamma_{MN}$, and the factor ${1}/{{2L}}$ in
front of the integral is introduced for convenience and chosen so
that the dimensions of the bulk gauge fields and coupling
constants be the same as in the four-dimensional theory, the field
strength tensors are given by
\begin{eqnarray}\label{F1}
&&W^{i}_{MN}=\partial_{M}W^{i}_{N}-\partial_{N}W^{i}_{M}+g\epsilon^{ikl}W^{k}_{M}W^{l}_{N},\\
\label{F1a} &&B_{MN}=\partial_{M}B_{N}-\partial_{N}B_{M},
\end{eqnarray}
and the fields satisfy the orbifold symmetry conditions
\begin{eqnarray*}
W^{i}_{\mu}(x,-y)&=&W^{i}_{\mu}(x,y),\quad
W^{i}_{4}(x,-y)=-W^{i}_{4}(x,y), \\
B_{\mu}(x,-y)&=&B_{\mu}(x,y),\quad B_{4}(x,-y)=-B_{4}(x,y).
\end{eqnarray*}

Next we will study  the excitations of the gauge bosons. To this
end we pass to the axial gauge, where the components $W^{i}_{4}$,
$B_{4}$ of the vector fields are equal to zero, and consider  only
the four-vector components of the five-dimensional gauge fields,
whose zero modes must play the role of the SM gauge bosons. From
action (\ref{action}) it is easy to get the following action for
the four-vector components of the five-dimensional gauge fields:
\begin{eqnarray}\label{effact}
S= \frac{1}{{2L}}\int
d^{4}xdy\left(-\frac{1}{4}\eta^{\mu\nu}\eta^{\alpha\beta}W^{i}_{\mu\alpha}W^{i}_{\nu\beta}+
e^{2\sigma}\frac{1}{2}\eta^{\mu\nu}\partial_{4}W^{i}_{\mu}\partial_{4}W^{i}_{\nu}
\right.\\
\nonumber \left.
-\frac{1}{4}\eta^{\mu\nu}\eta^{\alpha\beta}B_{\mu\alpha}B_{\nu\beta}+e^{2\sigma}\frac{1}{2}\eta^{\mu\nu}\partial_{4}B_{\mu}\partial_{4}B_{\nu}
\right).
\end{eqnarray}

Now, making the standard redefinition of the gauge fields
\begin{eqnarray}\label{gaugephys}
Z_{\mu}=\frac{1}{\sqrt{g^2+{g'}^2}}\left(gW^{3}_{\mu}-g'B_{\mu}\right),\,\,
A_{\mu}=\frac{1}{\sqrt{g^2+{g'}^2}}\left(g
B_{\mu}+g'W^{3}_{\mu}\right),\,\,
W^{\pm}_{\mu}=\frac{1}{\sqrt{2}}\left(W^{1}_{\mu}\mp
iW^{2}_{\mu}\right),
\end{eqnarray}
where $g$  and $g^\prime$  are the coupling constants of the gauge
groups $SU(2)$ and $U(1)$  respectively, we pass to the physical
degrees of freedom of the theory.

Let us rewrite  action (\ref{effact})  in terms of these redefined
fields and  decompose it into the quadratic part and the
interaction Lagragian. It takes the form
\begin{eqnarray}\label{effact2}\nonumber
S=\frac{1}{{2L}}\int
d^{4}xdy\left(-\frac{1}{2}\eta^{\mu\nu}\eta^{\alpha\beta}W^{+}_{\mu\alpha}W^{-}_{\nu\beta}
-\frac{1}{4}\eta^{\mu\nu}\eta^{\alpha\beta}A_{\mu\alpha}A_{\nu\beta}-\frac{1}{4}\eta^{\mu\nu}\eta^{\alpha\beta}Z_{\mu\alpha}Z_{\nu\beta}\right.\\\nonumber
 \left.
+\,
e^{2\sigma}\eta^{\mu\nu}\partial_{4}W^{+}_{\mu}\partial_{4}W^{-}_{\nu}+e^{2\sigma}\frac{1}{2}\eta^{\mu\nu}\partial_{4}A_{\mu}\partial_{4}A_{\nu}
+ e^{2\sigma}\frac{1}{2}\eta^{\mu\nu}\partial_{4}Z_{\mu}\partial_{4}Z_{\nu}\right.\\
 \left.+\,L_{WWV}+L_{WWVV}\right),\phantom{aaaaaaaaaaaaaaaaaaaaaaaaaaaaaaaaa}
\end{eqnarray}
where
$W^{\pm}_{\mu\nu}=\partial_{\mu}W^{\pm}_{\nu}-\partial_{\nu}W^{\pm}_{\mu}$,
$A_{\mu\nu}=\partial_{\mu}A_{\nu}-\partial_{\nu}A_{\mu}$,
$Z_{\mu\nu}=\partial_{\mu}Z_{\nu}-\partial_{\nu}Z_{\mu}$,
$L_{WWV}$ and  $L_{WWVV}$ are the gauge boson three-particle and
four-particle self interaction 5D Lagrangians respectively. The
 three-particle self interaction Lagrangian is explicitly given by
\begin{eqnarray}\label{triple_couplings}\nonumber
L_{WWV} &= & -i g\, \Bigl[\left(W_{\mu\nu}^+ W^{-\mu} -
W^{+\mu}W_{\mu\nu}^- \right)\left(A^\nu \sin\theta_W + Z^\nu
\cos\theta_W \right) \Bigr. \\
&+&\Bigl. W^{+}_{\mu}W_{\nu}^-\left(A^{\mu\nu} \sin\theta_W +
Z^{\mu\nu} \cos\theta_W \right) \Bigr],
\end{eqnarray}
and the four-particle self interaction  Lagrangian looks like
\begin{eqnarray}\label{quartic_couplings}\nonumber
L_{WWVV} &=& - \frac{g^2}{2}\, \Bigl[\left(W_{\mu}^+
W^{-\mu}\right)^2  - \left(W_{\mu}^+W^{+\mu}\right)\left(W_{\nu}^-
W^{-\nu}\right)\Bigr. \\ \nonumber &+&\Bigl.2\left(W_{\mu}^+
W^{-\mu}\right) \left(A_\nu \sin\theta_W + Z_\nu \cos\theta_W
\right) \left(A^\nu
\sin\theta_W + Z^\nu \cos\theta_W \right)\Bigr.\\
 &-&\Bigl.2 W_{\mu}^+ \left(A^\mu \sin\theta_W + Z^\mu \cos\theta_W \right)W^-_{\nu} \left(A^\nu
\sin\theta_W + Z^\nu \cos\theta_W \right)\Bigr].
\end{eqnarray}

In what follows, we assume that the bulk gauge symmetry
$SU(2)\times U(1)$ is spontaneously broken by the bulk Higgs field
(we will comment on the brane-localized Higgs field below). Here
we will not go into details of this mechanism \cite{Muck:2002af}.
We will just suppose that, as a result,  mass terms for the fields
$W_\mu$ and $Z_\mu$ are generated so that the masses of their zero
modes are given by the standard expressions
\begin{equation}\label{gmasses}
m_{W}=\frac{g  v}{2},\qquad m_{Z}=\frac{\sqrt{g^2+{g'}^{2}}\,
v}{2},
\end{equation}
$v$ denoting the  SM vacuum value of the Higgs field.

The equations for the wave functions $\chi_{V,n}$ and the masses
$m_{V,n}, \, V = A, W, Z,$ of the Kaluza-Klein modes can be
derived from action (\ref{effact2}) (here and below the subscript
$n$ denotes the number of the corresponding Kaluza-Klein mode).
When we  take into account the mass terms generated by spontaneous
symmetry breaking, they look like
\begin{eqnarray}
\label{Awf}
-m_{A,n}^{2}\chi_{A,n}-\partial_{4}(e^{2\sigma}\partial_{4}\chi_{A,n})=0,\\\label{Wwf}
-m_{W,n}^{2}\chi_{W,n}-\partial_{4}(e^{2\sigma}\partial_{4}\chi_{W,n})+m^2_{W}\chi_{W,n}=0,\\
\label{Zwf}
-m_{Z,n}^{2}\chi_{Z,n}-\partial_{4}(e^{2\sigma}\partial_{4}\chi_{Z,n})+m^2_{Z}\chi_{Z,n}=0.
\end{eqnarray}
As usually, we assume that the lowest (zero) Kaluza-Klein modes of
the 5D gauge fields coincide with the four-dimensional SM gauge
fields. It follows from  eq. (\ref{Awf}) that the solution for the
lowest mode of the field $A_{\mu}$ (the photon) is $m_{A,0}=0$ and
$\chi_{A,0}(y)\equiv \textrm{const} = 1$ (the latter equality is
due to our normalization of the bulk gauge fields), i.e. its wave
function does not depend on the coordinate of the extra dimension.
This  property of the solution guarantees the universality of the
electromagnetic charge \cite{Rubakov:2001kp}. The solutions of
eqs. (\ref{Wwf}) and (\ref{Zwf}) for the wave functions of the
lowest modes have the same property, if $m_{W,0}= m_{W}$ and
$m_{Z,0}= m_{Z}$. The only case, where the zero mode sector of a
five-dimensional model  exactly coincides with the electroweak
gauge boson sector of the SM, is the one, where the wave functions
$\chi_{W,0}(y)$ and $\chi_{Z,0}(y)$ do not depend on the
coordinate of the extra dimension.  However, to this end the
vacuum profile of the 5D Higgs field should be equal to
\begin{equation}\label{Higgs_vacuum}
v_0(y) = ve^{-\sigma(y)},
\end{equation}
i.e. it should be fine-tuned with background solution
(\ref{backgmetric}) for the metric \cite{Smolyakov:2015zsa}. In
this case the self-coupling constants of the massive gauge bosons
are defined in terms of the constants $g$ and $g'$ exactly in the
same way as in the ordinary SM. Also in this case the wave
functions $\chi_{V,n}, \, V = A, W, Z,$ of the gauge boson
excitations defined by eqs. (\ref{Awf}), (\ref{Wwf}), (\ref{Zwf})
are all equal and below will be denoted by $\chi_{n}(y)$. The
expansions of the 5D gauge fields in KK-modes look like
\begin{equation}\label{mode_decomp}
 V_\mu(x,y) = \sum_{n = 0}^{\infty}V_\mu^{(n)}(x) \chi_{n}(y), \, V = A, W,
 Z,
\end{equation}
and it is easy to check that the following relation holds for
these wave functions for an arbitrary KK-number $n > 0$ and an
arbitrary power $l > 0$ of the zero mode wave function
\begin{equation}\label{orthogonality}
\int_{-L}^L \chi_{n}(y)\left(\chi_{0}(y) \right)^l dy =
\int_{-L}^L \chi_{n}(y)\chi_{0}(y)  dy = 0.
\end{equation}

Below we  consider the case, where the masses of the zero modes of
the bulk fields $W_\mu$ and $Z_\mu$ are given by (\ref{gmasses})
and their wave functions are equal to unity due to our
normalization of the bulk gauge fields in action (\ref{action}).
In this case the masses of the first excitations $W^\prime$,
$Z^\prime$, and $\gamma^\prime$ are given by
\begin{eqnarray}\label{masses}
m_{W^\prime} &=&  m_{W,1}= \sqrt{m^2_{A,1} + m^2_{W}} \simeq
m_{\gamma^\prime} + \frac{m^2_{W}}{2m_{\gamma^\prime}}\\
m_{Z^\prime} &=&  m_{Z,1}= \sqrt{m^2_{A,1} + m^2_{Z}}\simeq
m_{\gamma^\prime} + \frac{m^2_{Z}}{2m_{\gamma^\prime}},
\end{eqnarray}
where $m_{A,1} = m_{\gamma^\prime}$ denotes the mass of
$\gamma^\prime$ and we have taken into account that
$m_{\gamma^\prime} \gg m_{W,Z}$. We emphasize that due to eq.
(\ref{orthogonality}) their wave functions $\chi_{1}(y)$ are
orthogonal to the wave functions of the gauge boson zero modes,
which are constants. To find the three-particle interactions of
the first gauge boson excitations with the SM gauge bosons, we
substitute the mode decompositions of the 5D gauge fields
$W_\mu(x,y)$, $Z_\mu(x,y)$ and $A_\mu(x,y)$ given by
(\ref{mode_decomp}) into interaction Lagrangian
(\ref{triple_couplings}), integrate with respect to the extra
dimension coordinate $y$ over the orbifold $S^1/Z_2$ and retain
only the terms with both  $W_\mu^\prime$, $Z_\mu^\prime$ or
$A_\mu^\prime$ and the SM gauge bosons. The resulting effective
three-particle interaction 4D Lagrangian of $W_\mu^\prime$,
$Z_\mu^\prime, A_\mu^\prime$ and the SM gauge bosons is given by
\begin{eqnarray}\label{triple_couplings_4D}\nonumber
L^{eff}_{WWV} = &-&i g\Bigl[\left(W_{\mu\nu}^{\prime+}
W^{\prime-\mu} - W^{\prime+\mu}W_{\mu\nu}^{\prime-}
\right)\left(A^\nu \sin\theta_W + Z^\nu \cos\theta_W \right)
\Bigr. \\\nonumber &+&\Bigl.
W^{\prime+}_{\mu}W_{\nu}^{\prime-}\left(A^{\mu\nu} \sin\theta_W +
Z^{\mu\nu} \cos\theta_W \right) \Bigr]\\\nonumber &-&i g
\Bigl[\left(W_{\mu\nu}^{\prime+} W^{-\mu} -
W^{\prime+\mu}W_{\mu\nu}^- \right)\left(A^{\prime\nu }\sin\theta_W
+ Z^{\prime\nu} \cos\theta_W \right) \Bigr. \\\nonumber &+&\Bigl.
W^{\prime+}_{\mu}W_{\nu}^-\left(A^{\prime\mu\nu} \sin\theta_W +
Z^{\prime\mu\nu} \cos\theta_W \right) \Bigr]\\\nonumber &-& ig
\Bigl[\left(W_{\mu\nu}^+ W^{\prime-\mu} -
W^{+\mu}W_{\mu\nu}^{\prime-} \right)\left(A^{\prime\nu}
\sin\theta_W + Z^{\prime\nu} \cos\theta_W \right) \Bigr.
\\ &+& \Bigl.
W^{+}_{\mu}W_{\nu}^{\prime-}\left(A^{\prime\mu\nu} \sin\theta_W +
Z^{\prime\mu\nu} \cos\theta_W \right) \Bigr].
\end{eqnarray}
Similarly we can find the effective  four-particle interaction 4D
Lagrangian (see Appendix).

Both these Lagrangians have the property that, due to
orthogonality condition (\ref{orthogonality}), a lowest excitation
of the SM gauge bosons cannot interact at tree level with two or
three SM gauge bosons. In particular, it means that the decays at
tree level of $W^{\prime}$, $Z^{\prime}$, and $\gamma^\prime$ into
two or three SM gauge bosons are forbidden. However, these bosons
can decay into SM fermions, because the wave functions of the zero
modes of the 5D fermions are not constant
\cite{Smolyakov:2015zsa}, and the corresponding coupling is
defined by the overlap integral of two fermion zero mode wave
functions and the wave function $\chi_{1}(y)$ of the first gauge
boson excitations. These overlap integrals are, of course, model
dependent and generally not equal to zero (see, e.g.,
\cite{Chang:1999nh,Gherghetta:2000qt,Djouadi:2007eg}). This
property also means that the lowest excitations of the SM gauge
bosons can decay into two or three SM gauge bosons via triangle or
box loop diagrams with SM fermions running in the loops, although
the decays are very much suppressed.

However, in the general case, where the vacuum solution for the 5D
Higgs field is not fine-tuned, the solutions for the wave
functions of the zero modes of the bulk fields $W_{\mu}$ and
$Z_{\mu}$, which correspond to the SM massive gauge bosons, are
not necessarily constant, and these decays can also take place due
to deviations of the zero mode gauge boson wave functions from
unity. In this case eq. (\ref{Awf}) remains the same, whereas eqs.
(\ref{Wwf}),(\ref{Zwf}) take the form:
\begin{eqnarray}
\label{Wwf1}
&-&\left(m_{W,n}^{2} - m^2_{W}\right)\chi_{W,n}-\partial_{4}(e^{2\sigma}\partial_{4}\chi_{W,n})+ \Delta M^2_{W}(y)\chi_{W,n}=0,\\
\label{Zwf1}
&-&\left(m_{Z,n}^{2} -
m^2_{Z}\right)\chi_{Z,n}-\partial_{4}(e^{2\sigma}\partial_{4}\chi_{Z,n})+\Delta
M^2_{Z}(y)\chi_{Z,n}=0,
\end{eqnarray}
where the extra terms $\Delta M^2_{V}(y),\,\, V = W,Z,$ depend on
the vacuum profile of the bulk Higgs field and result in
deviations  of the wave functions $\chi_{W,0}(y)$ and
$\chi_{Z,0}(y)$ from constant. This has the following
consequences.  In the five-dimensional theory under consideration
the  self-coupling of massive gauge bosons comes, as usually, from
the term $W^{i,\mu\nu}W^{i}_{\mu\nu}$, but now the corresponding
coupling constants are defined not only by the structure constants
of the SM gauge group, but also by the overlap integrals of the
wave functions $\chi_{W,0}(y)$ and $\chi_{Z,0}(y)$ over the space
of extra dimension. Moreover, in the general case, a modification
of the shapes of the zero mode gauge boson wave functions has an
influence on the electroweak observables, which was discussed in
detail in \cite{Csaki:2002gy,Burdman:2002gr}. In these papers, the
bulk gauge symmetry $SU(2)\times U(1)$ is spontaneously broken by
the brane-localized Higgs field. In this case the mass eigenstates
of the vector fields are always a mixture of their KK modes, and
the zero mode sector of the effective four-dimensional theory
cannot exactly reproduce the electroweak sector of the SM. In
particular, the wave functions of $W$- and $Z$-bosons are not
equal to unity, and the admixture of the wave functions of the
first excited mode is proportional to the squared ratio of the
masses of the SM gauge boson and its first excitation. These
deviations of the  wave functions of the massive SM gauge bosons
from unity lead to contributions to the electroweak precision
parameters, and the restrictions on their value put severe
constraints on the admissible masses of the first gauge boson
excitations. In paper \cite{Burdman:2002gr} it was shown that the
masses of the first gauge boson excitations should be larger than
$49\, TeV,$ which is well above the present day experimental lower
limit on their masses of the order of $5\, TeV$
\cite{Tanabashi:2018oca}.

However, in the case of the bulk Higgs field, the situation is
different, and there may exist gauge boson excitations with masses
below $49\, TeV$. The mixing of the gauge boson KK modes takes
place not due to the interaction with the vacuum wave function of
the bulk Higgs field, but due to the interaction with the
deviation of the vacuum wave function of the bulk Higgs field from
the vacuum wave function, which renders constant wave functions of
the SM gauge bosons. This deviation can be fairly small, and the
corrections to the masses and wave functions of the $W$- and
$Z$-bosons arising from the terms $\Delta M^2_{V}(y),\,\, V =
W,Z,$ can be small enough not to influence noticeably the
electroweak parameters. In this case we can use the standard
perturbation theory to find approximate solutions to eqs.
(\ref{Wwf1}),(\ref{Zwf1}). To first order of perturbation theory
the  masses of $W$- and $Z$-bosons and the masses of their first
excitations $W^\prime$ and $Z^\prime$ look like
\begin{eqnarray}\label{massW}
m_{W,0} &=&  \sqrt{ m^2_{W} + \left(\Delta M^2_{W}
\right)_{00}}\\
\label{massZ}
m_{Z,0} &=&  \sqrt{m^2_{Z} +
\left(\Delta M^2_{Z} \right)_{00}}\\
m_{W^\prime} &=&  m_{W,1}= \sqrt{m^2_{\gamma^\prime} + m^2_{W} +
\left(\Delta M^2_{W} \right)_{11}} \simeq
m_{\gamma^\prime} + \frac{m^2_{W} + \left(\Delta M^2_{W} \right)_{11}}{2m_{\gamma^\prime}}\\
m_{Z^\prime} &=&  m_{Z,1}= \sqrt{m^2_{\gamma^\prime} + m^2_{Z} +
\left(\Delta M^2_{Z} \right)_{11}}\simeq m_{\gamma^\prime} +
\frac{m^2_{Z}+ \left(\Delta M^2_{Z}
\right)_{11}}{2m_{\gamma^\prime}},
\end{eqnarray}
the wave functions of the lowest modes $\chi_{V,0}(y)$ and
$\chi_{V,1}(y) \,\,, V=W,Z,$ are given by
\begin{eqnarray}
\label{Wfp0} \chi_{V,0}(y)&=& \chi_{0}(y) - \frac{\left(\Delta
M^2_{V} \right)_{10}}{m_{\gamma^\prime}^{2}}\,\chi_{1}(y)
 - \sum_{n =2}^\infty \frac{\left(\Delta M^2_{V} \right)_{n0}}{m_{A,n}^{2}}\,\chi_{n}(y), \\
\label{Wfp1} \chi_{V,1}(y) &=& \chi_{1}(y) + \frac{\left(\Delta
M^2_{V} \right)_{01}}{m_{\gamma^\prime}^{2}}\,\chi_{1}(y) +
\sum_{n =2}^\infty \frac{\left(\Delta M^2_{V}
\right)_{n1}}{m_{\gamma^\prime}^{2} - m_{A,n}^{2}}\,\chi_{n}(y),
\end{eqnarray}
where the matrix elements $\left(\Delta M^2_{V} \right)_{mn} =
\left(\Delta M^2_{V} \right)_{nm}$ are defined as
$$
\left(\Delta M^2_{V} \right)_{mn} = \frac{1}{2L} \int_{-L}^L
\chi_{m}(y) \Delta M^2_{V}(y)  \chi_{n}(y) dy.
$$

Since $m_{W,0}$ and $m_{Z,0}$ should lie within the experimental
uncertainties $\Delta m_{W}, \,\Delta m_{Z}$  from the standard
masses $m_{W}$ and $m_{Z}$, the matrix elements of the
perturbations must satisfy the conditions
\begin{equation}\label{dmass}
|\left(\Delta M^2_{W} \right)_{00}| \simeq 2 m_{W} (\Delta m_{W})
\leq 2\, GeV^2, \quad |\left(\Delta M^2_{Z} \right)_{00}| \simeq 2
m_{Z} (\Delta m_{Z}) \leq 0.4\, GeV^2.
\end{equation}
However, in the model under consideration the perturbation terms
$\Delta M^2_{V}(y),\,\, V = W,Z,$ are proportional to the
difference of the squared wave function $v^2(y)$ of the bulk Higgs
field at hand and the squared  wave function $v^2_0(y)$
(\ref{Higgs_vacuum}) of the Higgs field, for which the wave
functions of the SM gauge bosons are constant, and we can use the
second estimate for both matrix elements.

If there exists a first excitation of the SM gauge bosons, the
deviation of its wave function from constant must be very small in
the model under consideration. In particular, the results of paper
\cite{Csaki:2002gy,Burdman:2002gr} imply that, in order not to
contradict the restrictions on the electroweak precision
parameters,  the admixture of the first KK mode in the wave
function of the massive SM gauge bosons, i.e. the coefficient in
front of $\chi_1(y)$ in eq. (\ref{Wfp0}), must satisfy the
condition
\begin{equation}\label{dWF}
\frac{|\left(\Delta M^2_{V} \right)_{10}|}{m_{\gamma^\prime}^{2}}
< 4\times 10^{-6}, \quad V=W,Z.
\end{equation}

The wave functions  $\chi_{V,0}(y)$ and $\chi_{V,1}(y), \, V=W,Z,$
are normalized to unity up to terms of second order in the
perturbations, which can be neglected. Due to the orthogonality of
the system of the unperturbed wave functions $\{\chi_{n}(y)\}$ the
calculation of the overlap integrals of one wave function
$\chi_{V,1}(y)$ (\ref{Wfp0})  with two or tree wave functions
$\chi_{V,0}(y)$ (\ref{Wfp1})  is very easy and gives the results
of the order $|\left(\Delta M^2_{V}
\right)_{10}|/m^2_{\gamma^\prime} \ll 1$, which is extremely small
due to (\ref{dWF}). This means that, in this case, the decays at
tree level of $W^\prime$, $Z^\prime$, and $\gamma^\prime$ to two
or three SM gauge bosons are very much suppressed. Below we
discuss this point in more detail.

The interaction Lagrangian that describes the decays of
$W^\prime$, $Z^\prime$, and $\gamma^\prime$ into two SM gauge
bosons looks like
\begin{eqnarray}\label{triple_couplings_A}\nonumber
L^{eff}_{WWV} = &-&i g \sin\theta_W I_W
\Bigl[\left(W_{\mu\nu}^{\prime+} W^{-\mu} + W_{\mu\nu}^+
W^{\prime-\mu} - W^{\prime+\mu}W_{\mu\nu}^{-} -
W^{+\mu}W_{\mu\nu}^{\prime-} \right)A^\nu  \Bigr.
\\\nonumber &+&\Bigl. \left(W^{\prime+}_{\mu}W_{\nu}^{-} +
W^{+}_{\mu}W_{\nu}^{\prime-} \right) A^{\mu\nu} \Bigr]\\
\nonumber  &-&i g \cos\theta_W I_{WZ}
\Bigl[\left(W_{\mu\nu}^{\prime+} W^{-\mu} + W_{\mu\nu}^+
W^{\prime-\mu} - W^{\prime+\mu}W_{\mu\nu}^{-} -
W^{+\mu}W_{\mu\nu}^{\prime-} \right)Z^\nu  \Bigr.
\\\nonumber &+&\Bigl. \left(W^{\prime+}_{\mu}W_{\nu}^{-} +
W^{+}_{\mu}W_{\nu}^{\prime-} \right) Z^{\mu\nu} \Bigr]\\
\nonumber &-&i g I_{WW} \Bigl[\left(W_{\mu\nu}^{+} W^{-\mu} -
W^{+\mu}W_{\mu\nu}^- \right)\left(A^{\prime\nu }\sin\theta_W +
Z^{\prime\nu} \cos\theta_W \right) \Bigr. \\ &+&\Bigl.
W^{+}_{\mu}W_{\nu}^-\left(A^{\prime\mu\nu} \sin\theta_W +
Z^{\prime\mu\nu} \cos\theta_W \right) \Bigr],
\end{eqnarray}
where the overlap integrals are approximately given by
\begin{equation}\label{overlaps}
I_W = -\frac{(\Delta M^2_{W})_{10}}{m^2_{\gamma^\prime}}, \quad
I_{WZ} = -\frac{(\Delta M^2_{W})_{10} + (\Delta
M^2_{Z})_{10}}{m^2_{\gamma^\prime}}, \quad I_{WW} =
-\frac{2(\Delta M^2_{W})_{10}}{m^2_{\gamma^\prime}}.
\end{equation}

First, let us consider the decays of $W^\prime$ with mass $6\,
TeV,$ which is close to the current experimental limit
\cite{Tanabashi:2018oca}. It has two two-body decay modes into the
SM gauge bosons $W,Z$ and $W,\gamma$. The decay of $W^\prime$ to
the massive bosons $W$ and $Z$ dominates due to the contributions
of the longitudinal modes. Its partial width can be calculated
from Lagrangian (\ref{triple_couplings_A}) and in the leading
approximation looks like
\begin{eqnarray}\label{width_Wp}
\Gamma(W^{\prime +} \rightarrow W^{+} Z) &\simeq& \frac{g^2
\cos^2\theta_W I^2_{WZ} m^5_{W^\prime}}{192 \pi m^2_{W} m^2_{Z}}\\
\label{width_Wp1}  &<&\frac{\alpha \cot^2\theta_W }{48 } \,
\frac{4 |\left(\Delta M^2_{Z} \right)_{10}|^2  m^5_{W^\prime}}{m^2_{W} m^2_{Z}m_{\gamma^\prime}^{4}}\\
 &\simeq& \frac{\alpha \cot^2\theta_W  }{48}\,\, 56\, GeV \simeq 28\, MeV,
\end{eqnarray}
where equation (\ref{width_Wp1}) has been obtained from equation
(\ref{width_Wp}) with the help of (\ref{overlaps}) and the last
ratio in (\ref{width_Wp1})  has been calculated using eq.
(\ref{dWF}). Thus, for a mass of $W^\prime$ close to the current
experimental limit \cite{Tanabashi:2018oca} the width is less than
$28\, MeV$. The decay of $W^\prime$  to $W$ and $\gamma$ is even
more suppressed. Similar reasonings show that the decays of
$Z^\prime$ and $\gamma^\prime$ to two $W$-bosons are very much
suppressed just like those of $W^\prime$.

However, the decays of the first SM gauge boson excitations to the
SM fermions turn out to be unsuppressed. Let us again consider the
decays of $W^\prime$. In the leading order the decay width of
$W^\prime$ into an SM fermion is given by
\begin{equation}\label{decay_f}
\Gamma(W^{\prime +} \rightarrow f_u\bar f_d) \simeq N_C \frac{g^2
I^2_{Wf} m_{W^\prime}}{48 \pi},
\end{equation}
where $f_u(f_d)$ denote upper (lower) particles of the SM fermion
doublets, $N_C = 3(1)$ is the color factor for quarks (leptons),
and $I_{Wf}$ stands for the overlap integral. In papers
\cite{Chang:1999nh,Gherghetta:2000qt} the latter was estimated to
be of the order $I_{Wf} \sim 4$ in the unstabilized
Randall-Sundrum model. In the case of stabilized Randall-Sundrum
model one can conservatively estimate it to be of the order of
unity. Then eq. (\ref{decay_f}) gives the partial decay widths of
$W^\prime$ into SM fermions of the order of $1.5\, GeV$ for the
excitation mass of the order of $6\, TeV$. Obviously, similar
estimates can be obtained for the partial decay widths into pairs
of the SM fermions of  $Z^\prime$ and $\gamma^\prime$.

It is necessary to point out that there are also three-body decays
of the first gauge boson excitations such as $W^{\prime +}
\rightarrow W^{+} Z Z$. The interaction Lagrangian describing
these decays can be obtained from the four-particle interaction
Lagrangian in the Appendix by replacing one field of a gauge boson
excitation by the field of the corresponding SM gauge boson in
each term. The overlap integral factors in this Lagrangian are
similar to those in eq. (\ref{overlaps}) and are of the same order
of magnitude. Therefore, the corresponding three-body decay widths
are suppressed by the factor $g^2$ and three-body decay phase
space and enhanced by the ratio ${m^2_{\gamma^\prime}}/{m^2_{W}}$
compared to the two-body decay case. Numerically, the three- and
two-body decay widths are roughly of  the same order for
$W^\prime$ mass about $6\, TeV$. However, for larger masses the
three-body decays begin to dominate the two-body decays to gauge
bosons, nevertheless their widths remaining much smaller than
those of  the decays to fermions.

Thus, we see that, in the brane-world models with bulk Higgs
field, the decays of the first gauge boson excitations into the SM
gauge bosons are very much suppressed compared to their decays
into the SM fermions for the excitation masses close to the
current experimental limit. However, we believe that, due to
(\ref{dmass}), where the matrix elements of the perturbation
operator are very small, estimate (\ref{dWF}) is very
conservative, and in reality  the decays of the first gauge boson
excitations into the SM gauge bosons can be very much suppressed
for larger masses of the gauge boson excitations as well.

\section{Conclusion}
In the present paper we have studied the interactions of the
electroweak  gauge boson  excitations in models with warped extra
dimensions and the SM fields propagating in the bulk. It has been
found that they are rather different from the interaction
properties of these excitations in other models. In particular, we
have shown that the couplings of the lowest gauge boson
excitations $W^{\prime}$, $Z^{\prime}$, and $\gamma^\prime$ to the
SM gauge bosons treated as the zero modes of the 5D gauge fields
are either exactly equal to zero or very much suppressed. In the
former case, we have explicitly found the three-particle and
four-particle interaction Lagrangians of the gauge boson
excitations $W^{\prime}$, $Z^{\prime}$, and $\gamma^\prime$ and
the SM gauge bosons. At the same time the couplings of
$W^{\prime}$, $Z^{\prime}$, and $\gamma^\prime$ to the SM fermions
are non-zero allowing for their production and decays. These
properties of the gauge boson excitations in models with warped
extra dimensions and the SM fields, including the Higgs field,
propagating in the multidimensional bulk distinguish them from the
gauge boson excitations in other models beyond the SM. For
example, in models with the brane-localized Higgs field the KK
modes of  the gauge fields are always coupled, and the couplings
to the longitudinal modes of the SM gauge fields are proportional
to the squared masses of the SM gauge bosons leading to an
enhancement of the corresponding decay rates \cite{Agashe:2007ki}.
Such a scenario leads to admissible masses of the first
excitations above $49\, TeV$ \cite{Burdman:2002gr}. In the models
with an extension of the SM gauge group \cite{Langacker:1989xa}
the EW gauge boson excitations can couple to both the SM gauge
bosons and fermions with approximately the same strength. Even in
the UED models with flat extra dimension
\cite{Appelquist:2000nn,Rizzo:2001sd,Macesanu:2002ew,Belanger:2012mc}
the couplings of the excitations to both the SM gauge bosons and
fermions are different from the considered case: they are both
very much suppressed, and the excitations are expected to be
long-lived particles. Thus, the interactions of the electroweak
gauge boson excitations are rather different in different
extensions of the SM and, if extra vector bosons are found at the
LHC, their interaction properties may  point out a theory beyond
the SM, to which they may  belong.

Finally, we would like to note that the discussed property of
vanishing or strongly suppressed  couplings of the first excited
KK modes of the electroweak gauge bosons to the SM gauge bosons
can be important not only for searching and interpreting the
signals at the LHC, but also for analyzing dark matter scenarios
with vector mediators arising in  models with extra dimensions.

\section*{Acknowledgements}
The authors are grateful to  S.~Keizerov,  E.~Rakhmetov, and
M.~Smolyakov  for interesting discussions and useful remarks. The
reported study was funded by RFBR and CNRS, project number
20-52-15005.

\newpage
\section*{Appendix}
Substituting the mode decompositions of the 5D gauge fields
$W_\mu(x,y)$, $Z_\mu(x,y)$ and $A_\mu(x,y)$ (\ref{mode_decomp})
into interaction 5D Lagrangian (\ref{quartic_couplings}),
integrating with respect to the extra dimension coordinate $y$
over the orbifold $S^1/Z_2$ and retaining only the terms with both
$W_\mu^\prime$, $Z_\mu^\prime$ or $A_\mu^\prime$ and the SM gauge
bosons, we get the following  effective three-particle interaction
4D Lagrangian of $W_\mu^\prime$, $Z_\mu^\prime, A_\mu^\prime$ and
the SM gauge bosons:

\begin{eqnarray*}\label{quartic_couplings_4D}\nonumber
L_{WWVV} = -& {\frac{g^2}{2}}&\Bigl[\left(W_{\mu}^{\prime+}
W^{-\mu}\right)\left(W_{\nu}^{\prime+} W^{-\nu}\right) +
\left(W_{\mu}^+ W^{\prime-\mu}\right)\left(W_{\nu}^+
W^{\prime-\nu}\right)\Bigr.
\\\nonumber &+&\Bigl. 2 \left(W_{\mu}^{\prime+}
W^{\prime-\mu}\right)\left(W_{\nu}^+ W^{-\nu}\right) -
2\left(W_{\mu}^+W^{\prime+\mu}\right)\left(W_{\nu}^-
W^{\prime-\nu}\right)\Bigr. \\ \nonumber & -&\Bigl.
\left(W_{\mu}^{\prime+} W^{\prime+\mu}\right)\left(W_{\nu}^-
W^{-\nu}\right) -
\left(W_{\mu}^+W^{+\mu}\right)\left(W_{\nu}^{\prime-}
W^{\prime-\nu}\right) \Bigr. \\ \nonumber
&+&\Bigl.2\left(W_{\mu}^{\prime+} W^{\prime-\mu}\right)
\left(A_\nu \sin\theta_W + Z_\nu \cos\theta_W \right) \left(A^\nu
\sin\theta_W + Z^\nu \cos\theta_W \right)\Bigr.\\
\nonumber &+&\Bigl.4\left(W_{\mu}^{\prime+} W^{-\mu}\right)
\left(A^\prime_\nu \sin\theta_W + Z^\prime_\nu \cos\theta_W
\right) \left(A^\nu
\sin\theta_W + Z^\nu \cos\theta_W \right)\Bigr.\\
\nonumber &+&\Bigl.4\left(W_{\mu}^+ W^{\prime-\mu}\right)
\left(A^\prime_\nu \sin\theta_W + Z^\prime_\nu \cos\theta_W
\right) \left(A^\nu \sin\theta_W + Z^\nu \cos\theta_W
\right)\Bigr.\\\nonumber &+&\Bigl.2\left(W_{\mu}^{+}
W^{-\mu}\right) \left(A^\prime_\nu \sin\theta_W + Z^\prime_\nu
\cos\theta_W \right) \left(A^{\prime\nu} \sin\theta_W + Z^{\prime\nu} \cos\theta_W \right)\Bigr.\\
\nonumber &-&\Bigl.2 W_{\mu}^{\prime+} \left(A^{\prime\mu}
\sin\theta_W + Z^{\prime\mu} \cos\theta_W \right)W^-_{\nu}
\left(A^\nu \sin\theta_W + Z^\nu \cos\theta_W \right)\Bigr.\\
\nonumber &-&\Bigl.2 W_{\mu}^+ \left(A^\mu \sin\theta_W + Z^\mu
\cos\theta_W \right)W^{\prime-}_{\nu} \left(A^{\prime\nu}
\sin\theta_W + Z^{\prime\nu} \cos\theta_W \right)\Bigr.\\
\nonumber &-&\Bigl.2 W_{\mu}^{\prime+} \left(A^\mu \sin\theta_W +
Z^\mu \cos\theta_W \right)W^{\prime-}_{\nu} \left(A^\nu
\sin\theta_W + Z^\nu \cos\theta_W \right)\Bigr.\\
\nonumber &-&\Bigl.2 W_{\mu}^+ \left(A^{\prime\mu} \sin\theta_W +
Z^{\prime\mu} \cos\theta_W \right)W^{-}_{\nu} \left(A^{\prime\nu}
\sin\theta_W + Z^{\prime\nu} \cos\theta_W \right)\Bigr.\\
\nonumber &-&\Bigl.2 W_{\mu}^{\prime+} \left(A^\mu \sin\theta_W +
Z^\mu \cos\theta_W \right)W^-_{\nu} \left(A^{\prime\nu}
\sin\theta_W + Z^{\prime\nu} \cos\theta_W \right)\Bigr.\\\nonumber
&-&\Bigl.2 W_{\mu}^+ \left(A^{\prime\mu} \sin\theta_W +
Z^{\prime\mu} \cos\theta_W \right)W^{\prime-}_{\nu} \left(A^\nu
\sin\theta_W + Z^\nu \cos\theta_W \right)\Bigr]\\ \nonumber +&
{g_{eff}}&\Bigl[2\left(W_{\mu}^{\prime+}
W^{\prime-\mu}\right)\left(W_{\nu}^{\prime+} W^{-\nu}\right) +
2\left(W_{\mu}^{\prime+} W^{\prime-\mu}\right)\left(W_{\nu}^+
W^{\prime-\nu}\right)\Bigr.\\\nonumber &-&\Bigl.
2\left(W_{\mu}^{\prime+}
W^{\prime+\mu}\right)\left(W_{\nu}^{\prime-} W^{-\nu}\right) -
2\left(W_{\mu}^{\prime+} W^{+\mu}\right)\left(W_{\nu}^{\prime-}
W^{\prime-\nu}\right)\Bigr.\\
\nonumber &+&\Bigl.6\left(W_{\mu}^{\prime+} W^{\prime-\mu}\right)
\left(A^\prime_\nu \sin\theta_W + Z^\prime_\nu \cos\theta_W
\right) \left(A^\nu
\sin\theta_W + Z^\nu \cos\theta_W \right)\Bigr.\\
\nonumber &+&\Bigl.3\left(W_{\mu}^{\prime+} W^{-\mu}\right)
\left(A^\prime_\nu \sin\theta_W + Z^\prime_\nu \cos\theta_W
\right) \left(A^{\prime\nu} \sin\theta_W + Z^{\prime\nu}
\cos\theta_W \right)\Bigr.\\ \nonumber
 &+&\Bigl.3\left(W_{\mu}^{+} W^{\prime-\mu}\right)
\left(A^\prime_\nu \sin\theta_W + Z^\prime_\nu \cos\theta_W
\right) \left(A^{\prime\nu} \sin\theta_W + Z^{\prime\nu}
\cos\theta_W \right) \\\Bigr.
 &-&\Bigl.3 W_{\mu}^{\prime+} \left(A^{\prime\mu} \sin\theta_W + Z^{\prime\mu} \cos\theta_W \right)W^{\prime-}_{\nu} \left(A^\nu
\sin\theta_W + Z^\nu \cos\theta_W \right)\Bigr.\\
\nonumber &-&\Bigl. 3 W_{\mu}^{\prime+} \left(A^\mu \sin\theta_W +
Z^\mu \cos\theta_W \right)W^{\prime-}_{\nu}
 \left(A^{\prime\nu}
\sin\theta_W + Z^{\prime\nu} \cos\theta_W \right)\Bigr.\\
\nonumber
 &-&\Bigl.3 W_{\mu}^{\prime+} \left(A^{\prime\mu} \sin\theta_W + Z^{\prime\mu} \cos\theta_W \right)W^{-}_{\nu} \left(A^{\prime\nu}
\sin\theta_W + Z^{\prime\nu} \cos\theta_W \right)\Bigr.\\
&-&\Bigl.3 W_{\mu}^{+} \left(A^{\prime\mu} \sin\theta_W +
Z^{\prime\mu} \cos\theta_W \right)W^{\prime-}_{\nu}
\left(A^{\prime\nu} \sin\theta_W + Z^{\prime\nu} \cos\theta_W
\right)\Bigr],
\end{eqnarray*}
where
$$
{g_{eff}} =   \frac{g^2}{{2L}} \int_{-L}^L
\left(\chi_1(y)\right)^3 dy.
$$

\end{document}